\documentclass[showpacs,prb,8pt,twocolumn,]{revtex4}
\usepackage{graphicx}

\begin{document}

\title{Weak Ferromagnetism Accompanying Loop current order in  Underdoped Cuprates}
\author{Vivek Aji, Arkady Shekhter and C. M. Varma}
\address{Department of Physics and Astronomy, University of California,
Riverside, CA 92521} 
\begin{abstract}

We discuss the necessary symmetry conditions, and the different ways in which they can be physically realized, for the occurrence of ferromagnetism accompanying the loop-current orbital magnetic order observed by polarized neutron diffraction experiments (or indeed any other conceivable principal order) in the underdoped phase of cuprates. We contrast the Kerr effect experiments in single crystals observing ferromagnetism with  
the direct magnetization measurements in large powder samples, which do not observe it. We also suggest experiments to resolve the differences among the experiments, all of whom we believe to be correct. 
\end{abstract}

\maketitle

\noindent{\it Introduction}

\begin{figure}
  \includegraphics[width=0.7\columnwidth]{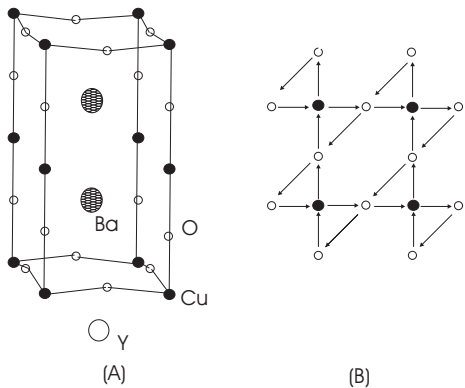}
  \caption{(A)Crystal structure of YBCO and (B) the Current pattern in the observed time reversal violating states}
  \label{cp}
\end{figure}

Polarized neutron scattering experiments \cite{FAQ,mook,greven} and dichroic Angle Resolved Photoemission experiments \cite{AK} have revealed an unusual ordered magnetic phase in underdoped cuprates. The transition temperature of this phase $T_g(x)$ varies systematically with doping $x$, consistent with the ``pseudogap'' temperature at which a characteristic change in transport and thermodynamic properties is observed. The symmetry of this phase is consistent with a loop-current phase predicted for the ``pseudogap'' phase \cite{CMV}. This phase has two orbital moments per unit cell with opposite mutual orientation, (see Fig.~\ref{cp}). The magnitude of the magnetic moment at low temperatures is $O(10^{-1}) \mu_B$ in each triangular O-Cu-O plaquette for $YBa_2Cu_3O_{6.7}$ \cite{FAQ,mook}. Similar magnitude is also found in underdoped single layer Hg-cuprates \cite{greven}.

Recently high precision Kerr effect experiments\cite{kapitulnik} on underdoped cuprates have been performed with light incident normal to the Cu-O planes. A rotation of polarization is discovered in underdoped cuprates below a temperature $T_K(x)$, which vanishes near optimal doping and systematically increases as $x$ is decreased below this value. $T_K(x)$ follows the same trend as $T_g(x)$ defined above but the best fits to the data place it $30 \pm 10$ K below $T_g(x)$. A rotation of polarization of reflected light is the classic signature of ferromagnetic ordering. Given the wavelength of light and the direction of incidence, the experiment implies a ferromagnetic moment normal to the planes. From the magnitude of the rotation angle, the magnitude of the moment at low temperatures is estimated to be of order $10^{-4}- 10^{-5}\mu_{B}$ per unit cell, i.e. 3 to 4 orders of magnitude smaller than the magnetic moment of the ordered loop currents.

Given these experimental facts, it would  be unreasonable that the ferromagnetic order is an independent spontaneous symmetry breaking. Almost certainly, it is an accompaniment to the much larger loop-current magnetic order. This is mandated if there exists a term in the free energy which is a product of the loop-current order parameter and the ferromagnetic order.  However, given a perfect {\it centered} orthorhombic symmetry of  crystals of $YBa_2Cu_3O_{6.7}$, which has a center of inversion at the center of the orthorhombic cell between the two Ba-atoms in Fig.~\ref{cp}, such a term is not allowed. The loop-current order is odd under both time reversal $R$ and inversion $I$ but even in the product $RI$. The ferromagnetic order is odd in $R$ but even in $I$. Therefore a bilinear of the requisite kind is impossible unless $I$ is independently broken by a lattice or magnetic distortion. The purpose of this Letter is to discuss the various possibilities for this to happen, to discuss their microscopic content, and suggest ways to distinguish between them by experiments. In suggesting various possiblities, we also have in mind the recent high precision magnetization measurements in large (powder) samples of underdoped $YBa_2Cu_3O_{6+\delta}$, which observe a signature of loop current order but do not see hysteresis characteristic of ferromagnetism with magnetic moments of $O(10^{-6})\mu_{B}$ or larger per unit cell, at either $T_K(x)$ or $T_g(x)$.\\

One may also ask the general inverse question: Order parameter of what symmetry allows ferromagnetism as an accompanying order parameter in the bulk of a crystal of centered orthorhombic symmetry such as an ideal sample of $YBa_2Cu_3O_{6+\delta}$? Since 
Ferromagnetism with moments perpendicular to the planes breaks all  symmetries under reflections on all planes perpendicular to the two-dimensional planes and preserves reflection on the plane parallel to the two dimensional plane through the center of inversion, the only possibility are antiferromagnetic arrangement of  moments with directions in the plane and with opposite orientations in the two bilayers. Such arrangements would be relatively easy to find in neutron diffraction experiments since they break lattice translational symmetry and produce new Bragg spots. No such broken symmetry has been observed \cite{bourges-pc}.

\noindent {\it Orbital Magnetic order in Underdoped Cuprates}

The unit cell for YBCO and a representative current loop pattern in a unit cell are shown in Fig.~\ref{cp}. In Ref.~\onlinecite{aji2} we have shown that spin orbit scattering induces an in-plane spin magnetic order in the presence of the observed loop-current order with a moment of O($10^{-2} \mu_B)$. Such an order is consistent with the symmetry requirements for the Dzialoshinskii-Moriya (DM) interactions and is in fact mandated by symmetry. The generated spin-moments are in plane and they are oppositely directed in the two bilayers per unit cell so that there is no net moment in a unit cell.

\begin{figure}
  \includegraphics[width =0.5\columnwidth]{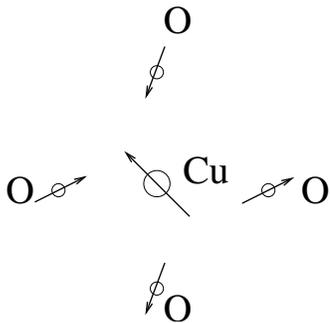}
  \caption{Schematic spin order in YBCO for the domain shown in the
  presence of exchange interaction. The spins need not be collinear and the angle
  between the net oxygen spins (summed over all oxygens in an unit cell) and the copper
  atom is dictated by the relative strength of the exchange and ordering fields.}
  \label{soj}
\end{figure}

When the exchange interaction between the copper and oxygen spins is also considered, a non-collinear spin-order on copper's and oxygens is possible which is in general not commensurate with the lattice. A schematic representation of the ordered spins is shown in Fig.~\ref{soj}. The neutron scattering experiments are not yet sensitive to enough to discard this possibility.

The in-plane  moments of Ref.~\onlinecite{aji2} are generated due to spin-orbit interaction and were motivated by the neutron diffraction results, which show a substantial angle of the magnetic moments with respect to the normal to the planes.  While the generation of such  moments through spin-orbit interactions is required by symmetry, their magnitude and therefore importance in understanding the neutron diffraction results is not definitely established. Another possibility for the in-plane component of the order (with the same lattice symmetry) are pyramidal current loops \cite{weber}, where the apical oxygens participate in the current loops. The general symmetry considerations that we use in this paper  are equally valid for such a contribution to the ordered magnetic moments.

The ferromagnetic moment perpendicular to the planes breaks time-reversal but preserves inversion, while the loop-current order and the in-plane spin-order in the absence of exchange interactions, are invariant under the product of time-reversal and inversion. For tetragonal {\it crystalline}  symmetry, the magnetic point group (in the presence of loop-current order) is ($m\underbar{m}m$) \cite{rb, simon, dv}. This has the following elements: the identity, twofold rotations around $\hat{y} (C_y)$ reflections in the $x-y$ and $y-z$ planes $(\sigma_z$ and $\sigma_x$), $RI, R\sigma_y, RC_x$ and $RC_z$. In orthorhombic {\it crystalline} symmetry, the magnetic point group is ($\underline{2}/m$), which has the symmetry elements identity, $\sigma_z$, $RC_z$ and $RI$. Neither of these groups allows ferromagnetism \cite{rb}.  However if inversion is also separately broken, the only point groups are ($m$), with elements identity and $\sigma_z$ or the triclinic ($1$) with no symmetries other than identity.  Both allow ferromagnetism.\\

\noindent{\it Possible Ways for a Ferromagnetic Order to accompany Loop Current Order}

We list the possible physical realizations of the symmetry breaking necessary for the ferromagnetic order:

(1)  A ferromagnetic order must accompany the loop-current order if the two triangular plaquettes with oppositely directed orbital moments in Fig.~\ref{cp}  have unequal areas and therefore have unequal magnetic moments. This cannot happen in the assumed crystal structure of $YBa_2Cu_3O_{6+\delta}$. But it follows if, for example, the crystal  loses the center of inversion by a relative movement of the Cu and O atoms in the metallic bilayers retaining the orthorhombic space-group symmetry. It also follows if the crystal has a triclinic distortion. This distortion need only be about $10^{-3}-10^{-4}$ of the lattice constant to get the requisite ferromagnetic moment.  Such weak distortions may not be measurable, even with present day experimental techniques. Moreover, the width of a Bragg spot in an experiment is limited in its sharpness by  the inverse of the typical linear dimension over which a crystal is perfect. So, the crystal would have to be nearly perfect over linear dimensions of order $10^{3}-10^{4}$ unit cells to tell the difference. This is not conceivable in the cuprates. It is important to mention that a ferromagnetic moment due to such distortions is not confined to being perpendicular to the Cu-O planes.

(2) Near the surface of the crystal, a polar-vector perpendicular to the surface naturally exists, breaking inversion. Its microscopic origin is surface relaxation of atoms. Such distortions decay in from the surface typically as the square of the distance. So, the ferromagnetic moment would also decay in similar manner. In the Kerr effect experiments light penetrates about $10^3$ Angstroms. Therefore the observed magnitude of the effect can only occur if the surface layer is distorted enough to produce a ferromagnetic moment of not much smaller than an order smaller than the loop-current moment measured in the bulk through polarized neutron scattering. The order of magnitude of the observed ferromagnetic order then can come from the top $10^2$ Angstroms. For this source of the ferromagnetic order also, the direction of the moment  need not be confined to perpendicular to the planes. We discuss below why we favor this mechanism.

(3) For the case of   $YBa_2Cu_3O_{6+\delta}$, as discussed below, the chain-oxygen  disorder or incommensurate order at general $\delta$  fulfills the symmetry requirements for ferromagnetism.

(4) The possible incommensurate in-plane spin-magnetic order shown in Fig.~\ref{soj} also lifts the center of inversion so as to permit ferromagnetism.

\noindent{\it Specific Examples}:

We now consider a few crystal structures of cuprates to find the conditions for such order.

$Y_1Ba_2Cu_3O_{6+x}$: The given crystal structure is orthorhombic with two $Cu-O_2$ layers per unit cell with a center of inversion between them. Therefore ferromagnetism is not possible unless some other lattice or spin distortion also occurs. As discussed an undetectably small deviation from the centered-orthorhombic symmetry is enough to produce the magnitude of ferromagnetism detected. The same is true for lifting of the center of inversion at the surface provided it is $O(10^{-1})$ and can be detected by low energy electron diffraction (LEED)  or other surface diffraction techniques. Other possibilities for the requisite symmetry
lowering to get ferromagnetic moment  are

(i) At ($x=0$) the chains are completely empty of oxygen and at  ($x=1$) they are completely full. Then the point of inversion in the crystal structure in between the Cu atoms in each bilayer is preserved. Therefore ferromagnetism is not possible at $x=0$ or $x=1$. For general, $0<x<1$, the oxygens in the chain are disordered or in specially prepared samples, they are ordered for some discrete values of $x$ in a manner incommensurate with the underlying orthorhombic lattice. The inversion symmetry in either case is lifted and ferromagnetism is possible in principle. (ii) Another possibility to satisfy D, as discussed in Ref.~\onlinecite{aji2} and in connection with Fig.~\ref{soj}, is that the in-plane ordered spin moments due to the leading spin-order effect are expected in general to be incommensurate with the lattice due to exchange effects lifting the inversion symmetry.

$La_{2-x}Sr_xCuO_4$ : The dimpling of the $Cu-O$ planes is such that there is no center of inversion in the unit cell. Ferromagnetism is then allowed. It ought to be mentioned that polarized
neutron diffraction finds spin flip intensity which suggests order with much smaller correlation length compared to YBCO but of the same integrated weight \cite{bourges-pc}.

$YBa_2Cu_4O_8$: The two chains in this orthorhombic crystal
structure are ordered. The only possibility for "weak
ferromagnetism"  is if there is an incommensurate in-plane spin
structure or through bulk or surface distortions discussed in the Introduction.

$BISCCO$: It is still not definitely determined whether the crystal structure does or does not have a center of inversion \cite{yam,pet}. In the former case the conditions are the same as in
$YBa_2Cu_4O_8$.\\

\noindent {\it Discussion: Hysteresis and Difference of $T_g(x)$ and $T_K(x)$}

In Ref.~\onlinecite{kapitulnik}, an interesting history dependence of the magnetic field history has been discovered. Usually the experiment is done by cooling in a field of order 1 Tesla, removing the field at very low temperatures and measuring finite dichroism of a specific sign as a function of increasing temperature, whose magnitude decreases to zero within the experimental resolution at a temperature $T_K(x)$. When the cooling field is reversed, the sign of dichroism reverses. All this is normal enough. But the experimentalists have also found that if they apply a field at $T=300K$, well above $T_K(x)$ for a few minutes and remove it, cool the sample in zero field and measure dichroism while increasing T, the same effect as above is obtained. The authors speculate that  ferrromagnetism with a moment $\lesssim 10^{-5} \mu_B$, their experimental resolution,  exists in the sample well above $T_K(x)$, which is aligned by the field and which in turn aligns the larger ferromagnetic moment observed below $T_K(x)$. We concur and would like to suggest that a tiny fraction of a  magnetic impurity distributed in the crystal but in a clustered form so that they are ordered locally but in mutually random direction would give this behavior. In a sufficiently large external field, such clusters orient along the field and remain so as the field is
removed at temperatures below their ordering. Suppose there is a linear local coupling between a defect moment $M_d$ at ${\bf r}_d$ and the ``intrinsic'' ferromagnetic moment $M({\bf r}_d)$ with a coupling constant $\lambda$. The ``intrinsic  moment'' at a site ${\bf r}$, then follows

\begin{equation}
\xi^2(T) \nabla^2 M({\bf r})= \lambda M_d({\bf r}_d).
\end{equation}
$\xi(T)$ is the intrinsic correlation length. Therefore the intrinsic moment grows in a region of linear dimension $\xi(T)$ around any  defect cluster. $\xi(T)$ grows as temperature is increased towards the intrinsic ordering temperature. When $\xi(T)$ reaches the order of the distance between the aligned magnetic defects, long-range order ensues. There is independent evidence through very precise magnetization measurements \cite{leridon} of ferromagnetic order at temperatures well above room temperature with average moment of $O(10^{-6})\mu_B$ per unit cell, presumed to be due to tiny clusters of $Fe_3O_4$ with concentration of $O(10^{-6})$. These measurements are however not done in the same samples as the Kerr effect measurements.

We now discuss how $T_K(x)$ may be systematically below but parallel to $T_g(x)$ by about $30-40 K$. A possibility, supported by the fact that this difference is more obvious in the single crystals in which the oxygen chains are ordered than in films, is that the pseudogap temperature of the ordered oxygen chain crystals is systematically lower than that of the samples in which neutron scattering is measured and which do not have the chain order. This is not a very likely possibility because the discrepancy between $T_K(x)$ and $T_g(x)$ is apparent  when they are plotted against the supercondcuting transition temperatures $T_c(x)$ of the samples.  Nevertheless, other measurements determining the pseudogap temperature of the chain ordered crystals may be worthwhile. If this possibility can be discarded, the possible mechanisms  for ferromagnetism discussed above can be narrowed. In all the possibilities discussed above except that due to surface induced ferromagnetism, the loop-current order and ferromagnetic transition temperatures must be the same. In general the surface layers have a different charge densities than the bulk. (In $YBaCuO$, they are known to have a higher doping concentration \cite{broun}). Therefore both their loop-current ordering temperature and the ferromagnetic transition temperature are different from that of the bulk.

The possibility of  ferromagnetism through surface-induced inversion breaking is also suggested by the fact that accurate bulk magnetization measurements \cite{leridon} with a sensitivity of better than $10^{-6}\mu_B/unit-cell$, do not observe the hysteresis loops characteristic of onset of ferromagnetism either at $T_K(x)$ or $T_g(x)$. If only $O(10^{-2}-10^{-3})$ of the sample (at the surface) had ferromagnetism, it would have gone undetected. As mentioned above Kerr effect measurements penetrate only about $10^3 \AA$ in from the surface and would detect such ferromagnetism. To establish this possibility firmly, we suggest that magnetization measurements be done in the same crystals as the Kerr effect measurements.

In conclusion, we would like to  underscore the fantastic sensitivity and the importance of the Kerr effect measurements using the principle of the Sagnac interferometer \cite{kapitulnik} which have clearly observed a broken time-reversal symmetry accompanying the orders of magnitude larger loop-current magnetic order in the pseudogap phase. This technique allows an ordered ferromagnetic moment to be detected which in the perfect crystal is forbidden by symmetry. But the magnetic moments are so tiny that, as we have discussed, it is enough to lift the  symmetry restrictions by lattice distortions (of the size of order of nuclear dimensions ) which may themselves be much harder to directly detect or through lifting of symmetry requirements at the surface. We also wish to reiterate the symmetry argument made in the introduction that primary order parameters to which the ferromagnetism may be an accompaniment in the perfectly centered orthorhombic crystals are easy to discover and so far are not found.

\noindent{\it Acknowledgements}: We wish to thank Aharon Kapitulnik for acquainting us with his experimental results and Steve Kivelson for discussions.  Steve Kivelson has independently come to the same conclusions as presented here regarding the impossibility of producing the ferromagnetic order in orthorhombic crystals with a
center of inversion. We also wish to thank Brigitte Leridon and Philippe Monod for acquainting us with their magnetization measurements and Philippe Bourges about the neutron scattering results.


\begin{thebibliography}{99}

\bibitem{FAQ}
B.~Fauque, Y.~Sidis, V.~Hinkov, S.~Pailh\`es, C.~T. Lin, X.~Chaud, and
  P.~Bourges.
\newblock {\em Phys. Rev. Lett.}, 96:197001, 2006.

\bibitem{mook}
H.~Mook.
\newblock arXiv:0802.3620

\bibitem{greven}
Yuan Li, Victor Baledent, Neven Barisic, Philippe Bourges, Yongchan Cho, Benoit
  Fauque, Yvan Sidis, Guichuan Yu, Xudong Zhao, and Martin Greven.
\newblock arXiv:0805.2959

\bibitem{AK}
A.~Kaminski, S.~Rosenkranz, H.~M. Fretwell, J.~C. Campuzano, Z.~Li, H.~Raffy,
  W.~G. Cullen, H.~You, C.~G. Olson, C.~M. Varma, and H.~Hochst.
\newblock {\em Nature}, 416:610, 2002.

\bibitem{CMV}
C.M. Varma.
\newblock {\em Phys. Rev. B}, 55:14554, 1997.

\bibitem{kapitulnik}
Jing Xia, Elizabeth Schemm, G.~Deutscher, S.A. Kivelson, D.~A. Bonn, W.~N.
  Hardy, R.~Liang, W.~Siemons, G.~Koster, M.~M. Fejer, and A.~Kapitulnik.
\newblock {\em Cond-mat}, 0711.2494, 2008.


\bibitem{bourges-pc}
P.~Bourges.
\newblock Private communication.

\bibitem{aji2}
V.~Aji and C.M. Varma.
\newblock {\em Phys. Rev.B}, 75:224511, 2007.

\bibitem{weber}
C.~webber.
\newblock PhD thesis, Ecole Polytechnique Federale De Lausanne, 2007.

\bibitem{rb}
R.R. Birss.
\newblock {\em Symmetry and Magnetism}.
\newblock Wiley-Interscience Inc. New York, 1964.

\bibitem{simon}
M.E. Simon and C.M. Varma.
\newblock {\em Phys. Rev. Lett}, 49:1545, 1982.

\bibitem{dv}
S.~Di Matteo and C.M. Varma.
\newblock {\em Phys. Rev. B}, 67:134502, 2003.

\bibitem{yam}
A.~Yamamoto, M.~Onoda, E.Takayama-Muromachi, F.~Izumi, T.~Ishigaki, and
  H.~Asano.
\newblock {\em Phys. Rev. B}, 42:4228, 1990.

\bibitem{pet}
V.~Petricek, Y.~Gao, P.~Lee, and P.~Coppens.
\newblock {\em Phys. Rev. B}, 42:387, 1990.

\bibitem{leridon}
B.~Leridon, P.~Monod, and D.~Colson.
\newblock arXiv:0806.2128 and rivate communication.

\bibitem{broun}
D.M. Broun, W.A. Huttema, P.J. Turner, S.~�zcan, B.~Morgan, R.~Liang, W.N.
  Hardy, and D.~A. Bonn.
\newblock {\em Phys. Rev. Lett.}, 99:237003, 2007.

\end{thebibliography}
\end{document}